\documentclass{article}

\usepackage[nonatbib,dblblindworkshop,preprint]{neurips_2026}
\workshoptitle{PhysWorldAI}

\usepackage[utf8]{inputenc} 
\usepackage[T1]{fontenc}    
\usepackage{hyperref}       
\usepackage{url}
\usepackage{hyperref}       
\usepackage{booktabs}       
\usepackage{amsfonts}       
\usepackage{nicefrac}       
\usepackage{siunitx}
\usepackage{microtype}      
\usepackage{xcolor}         

\usepackage{graphicx}
\usepackage{subfig}
\usepackage{pdflscape}

\usepackage[
  backend=biber,
  style=numeric-comp,   
  sorting=none,         
  natbib=true,          
  maxbibnames=99,       
  maxcitenames=2,       
  giveninits=true,
  url=false, doi=false, isbn=false,   
]{biblatex}
\usepackage{csquotes}

\usepackage[acronym,nomain]{glossaries-extra}%
\setabbreviationstyle[acronym]{long-short}

\newacronym{ai}{AI}{Artificial Intelligence}
\newacronym{doe}{DOE}{Department of Energy}

\newacronym{ae}{AE}{Alfven Eigenmode}
\newacronym{tm}{TM}{Tearing Mode}
\newacronym{elm}{ELM}{Edge Localized Modes}

\newacronym{ece}{ECE}{Electron Cyclotron Emission}
\newacronym{bes}{BES}{Beam Emission Spectroscopy}
\newacronym{mhr}{MHR}{Magnetics High Resolution}
\newacronym{co2}{CO$_2$}{CO$_2$ Interferometry}
\newacronym{cer}{CER}{Charge Exchange Recombination}
\newacronym{mse}{MSE}{Motional Stark Effect}
\newacronym{ts}{TS}{Thomson Scattering}
\newacronym{tangtv}{TangTV}{Tangential TV}
\newacronym{ech}{ECH}{Electron Cyclotron Heating}
\newacronym{nbi}{NBI}{Neutral Beam Injection}
\newacronym{rmp}{RMP}{Resonant Magnetic Perturbation}
\newacronym{arc}{ARC}{Affordable, Robust, Compact}
\newacronym{mast}{MAST}{Mega Amp Spherical Tokamak}
\newacronym{kstar}{KSTAR}{Korea Superconducting Tokamak Advanced Research}
\newcommand{\ddd}{DIII-D }

\newacronym{fsq}{FSQ}{Finite-Scalar-Quantization}
\newacronym{wfm}{WFM}{World Foundation Model}
\newacronym{rl}{RL}{Reinforcement Learning}
\newacronym{maskgit}{maskGIT}{Masked Generative Image Transformer}
\newacronym{ignite}{IGNITE}{Integrated and Generative Network for Interactive Tokamak Emulation}
\newacronym{gpu}{GPU}{Graphics Processing Unit}
\newacronym{hpc}{HPC}{High Performance Computing}

\title{IGNITE Tokamak World Model Architecture}

\author{%
  Peter Steiner \quad Azarakhsh Jalalvand \quad Nathaniel Chen \quad Kouroche Bouchiat \\
  Department of Mechanical and Aerospace Engineering\\
  Princeton University\\
  Princeton, NJ \\
  \texttt{\{peter.steiner,azarakhsh.jalalvand,nathaniel,bouchiat\}@princeton.edu} \\
  \AND
  Ricardo Shousha \quad SangKyeun Kim \\
  Princeton Plasma Physics Laboratory\\
  Princeton, NJ \\
  \texttt{\{rshousha,skim2\}@pppl.gov} \\
  \AND
  Egemen Kolemen \\
  Princeton University, Princeton Plasma Physics Laboratory \\
  Princeton, NJ \\
  \texttt{ekolemen@princeton.edu} \\
}

\begin{document}

\maketitle

\begin{abstract}
    We introduce IGNITE, a generative \gls{wfm} for fusion plasma behavior simulation trained in a self-supervised manner from over a decade of unlabeled experimental data at the \ddd National Fusion Facility. The core of IGNITE is a dynamics model that can simulate \ddd discharges from a given set of actuator trajectories. These trajectories can be supplied or generated on-the-fly from a textual prompt or from desired experimental outcomes. The model architecture consists of several spatio-temporal tokenizers that embed the different input modalities, including time-series like spatio-temporal measurement data, image sequences, and high-resolution spectrograms, each of which collected at vastly different time scales. The backbone is composed of an auto-regressive dynamics model that has the capacity to predict entire \ddd discharges given initial latent plasma states and actuator trajectories over a theoretical infinite horizon. IGNITE paves the way towards efficient AI-driven experimental planning and world modeling for nuclear fusion.
\end{abstract}

\section{Introduction}

Generative \gls{ai} has rapidly become a revolutionary tool for accelerating advances across a broad range of research fields. Its emergence may be comparable to the widespread adoption of computers decades ago, which fundamentally reshaped scientific research and enabled entirely new ways of working and discovery. Recent advances in \gls{ai} can be attributed to the development of the transformer architecture \cite{Vaswani2017attention}, and the democratization of powerful modern \gls{gpu} hardware and large-scale \gls{hpc} clusters. Together, with the progression of self-supervised learning, these advances have enabled billion-parameter models to be trained on vast amounts of previously underutilized unlabeled data, overcoming one of the major limitations of conventional supervised learning: the need for large, carefully labeled datasets. This led to significant breakthroughs in many fields such as image and video processing \cite{Arnab2021ViViT,Dehghani2023scaling,Wang2023videomae,Bruce2024genie,Dosovitskiy2021an}, bioinformatic \cite{Guo2025Foundation}, and audio signal processing \cite{VanDenOord2016wavenet,Wang2017tacotron,Shen2018natural,Kong2020hifi,Langman2025spectral}.

Another area in which such models have not yet, but could lead to major breakthroughs is fusion energy \cite{Churchill2025ai}. Nuclear fusion offers humanity a potentially clean, sustainable, and safe energy source, and decades of research and experimentation led to the physics basis of fusion power plants such as the \gls{arc} reactor \cite{Hillesheim2026overview}, which is expected to start operations in the mid \num{2030}s with the plan to deliver \SI{400}{\MW} net electrical power to the grid. However, much experimentation is still required to find scalable, high-performance plasma regimes \cite{Lawson1957some,Wagner1982regime}. Running these experiments is expensive, requiring significant investment since it depends on competitive allocation and expensive hardware. Given the long history of recorded data for nuclear fusion, there is promising opportunity to learn full experiments using \gls{ai} models.

Fusion is a particularly well-suited domain for self-supervised techniques because plasma behavior is extraordinarily complex, involving nonlinear and multiscale dynamics, hence, difficult to model and predict comprehensively. Consequently, accurate prediction and control of fusion plasmas using purely model-based approaches remain major challenges. At the same time, a large number of experimental fusion devices have been built around the world, spanning different configurations, operational capabilities, and diverse sets of diagnostics. Decades of experiments on these machines have generated enormous volumes of data, much of which remains underexplored. This combination of data richness and physical complexity provides exactly the kind of environment in which large generative or foundation models, trained using self-supervised approaches, could have a revolutionary impact.

  
In this paper, we propose \gls{ignite}, a new generative \gls{ai} model designed to simulate fusion plasma behavior, trained on a large dataset of over \num{8000} discharges ($\approx$ \SI{12}{\hour} of data) collected from more than a decade of experiments. It is not only the first large-scale world model for fusion plasma, but also the first multi-modal approach, working with signals at different time scales ranging from \si{\Hz} level to \si{\MHz} level, and different modalities such as time-series in different sampling frequencies, spectrograms, and a tokenizer design that is extensible to other diagnostics types such as image sequences. It is also fully controllable on a frame-by-frame level by supplying actuator trajectories.

Similar to many research areas, \gls{ai} has had a strong impact on nuclear fusion research in the past years due to its strength in handling high dimensional, nonlinear data \cite{Anirudh2023review}. Several publications like \cite{Degrave2022magnetic,Seo2024avoiding,Wang2025learning} have shown the potential of using reinforcement learning for controlling nuclear fusion devices. A key point of these publications is that the deployed \gls{ai} models were not only successful in their accuracy, but that they were also deployed on real tokamaks to actively control or design experiments.

These models often rely on a subset of curated data, and work on a subset of diagnostics or target a subset of physics. To address this, works that show the potential of \gls{ai} models for enabling cross-diagnostic analysis \cite{Jalalvand2025multimodal} and robustness against sensor failures during nuclear fusion experiments \cite{Shousha2023machine} suggest the possibility of a single model that is capable of working on multiple of these tasks in parallel like predicting several profiles in parallel. A recent wave of large-scale self-supervised models for nuclear fusion have been demonstrated this move towards a developing such a single model: A collaboration between \gls{mast} tokamak and IBM has proposed the multi-modal foundation model TokaMind \cite{Boschi2026tokamind}. It is an efficient transformer-based encoder (9.3M parameters) that maps a context window of MAST diagnostics to a target window. The context window consists of time series, profiles, and image sequences. The model was pre-trained with a masked reconstruction task of a \SI{50}{\ms} window of \gls{mast} diagnostics. A key point is that the model uses a training-free signal embedding per diagnostic, allowing for different sampling frequencies (spanning \SI{200}{\Hz} to \SI{500}{\kHz}). After pre-training, the model was fine-tuned on several downstream tasks from reconstruction to prediction. In the latter case, the prediction horizon was up to \SI{100}{\ms}, meaning two windows of each \SI{50}{\ms} duration. It is worth noting that the prediction task was always a window-to-window regression rather than autoregressive forecasting.

In another study  PanoMHD \cite{Noh2026panomhd} is developed as a multi-modal model for predicting plasma dynamics for control purposes at \gls{kstar}. This model with 457M parameters predicts the next step magnetic-fluctuation spectrograms together with the scalar plasma performance, based on historic magnetic fluctuations, plasma performance, actuators, and plasma shape parameters. Like TokaMind, PanoMHD is optimized to only predict the next window of \SI{50}{\ms}.  FusionMAE is another example of such efforts which applies non-causal masked autoencoder reconstruction to actuator and diagnostic channels of the HL-3 tokamak. This 14M parameter model is capable of restoring lost signals and imputing compressed signals ranging from plasma current $I_p$ to lower hybrid resonant heating $P_{LH}$ \cite{Yang2026fusionmae}.

While TokaMind, PanoMHD, and FusionMAE establish that self-supervised pretraining on heterogeneous tokamak diagnostics yields transferable representations, they are formulated as conditioned predictors or reconstructors over a single short window. FusionMAE compresses a \SI{10}{\ms} slice into an embedding and imputes missing channels, with no temporal-dynamics model, its authors noting that capturing dynamics would require an autoregressive reformulation. TokaMind maps a context window to a target window ($\leq\SI{100}{\ms}$). PanoMHD predicts the next \SI{50}{\ms} window of magnetic-fluctuation spectrograms from a short history. FusionMAE predicts 10ms given a consistent input sample-grid of \SI{1}{\kHz}.

\gls{ignite} builds on these ideas with methods from state-of-the-art spectrogram \cite{Langman2025spectral}, video \cite{Arnab2021ViViT,Gupta2023maskvit,Villegas2023phenaki}, and time-series models \cite{Kong2020hifi,Langman2025spectral} as well as world models like Genie \cite{Bruce2024genie}. Similar to Genie, we use spatiotemporal transformers \cite{Xu2021spatial} in our dynamic model. We use \gls{fsq} encoders \cite{Mentzer2024finite} to tokenize the diagnostics and actuators, and these tokens are passed to a dynamic model that autoregressively predicts the tokens for the next frame.

\section{The IGNITE world model}

\gls{ignite} is a generative multi-modal actuator-conditioned world model of the \ddd tokamak. It takes a sequence of \num{20} windows of \SI{50}{\ms} duration of heterogeneous diagnostics plus seven actuator groups and autoregressively forecasts the diagnostics in the subsequent windows. Each modality is tokenized independently, and we use discrete \gls{fsq} codecs for all modalities. The world model thus learns dynamics in a fixed latent space. The architecture of IGNITE is adapted from the Genie architecture \cite{Bruce2024genie} and thus based on the transformer architecture \cite{Vaswani2017attention}. We also use the memory-efficient spatiotemporal transformer architecture from \cite{Xu2021spatial} to balance the model capacity with computational constraints.

\begin{figure}[!htb]
    \centering
    \includegraphics[width=\textwidth]{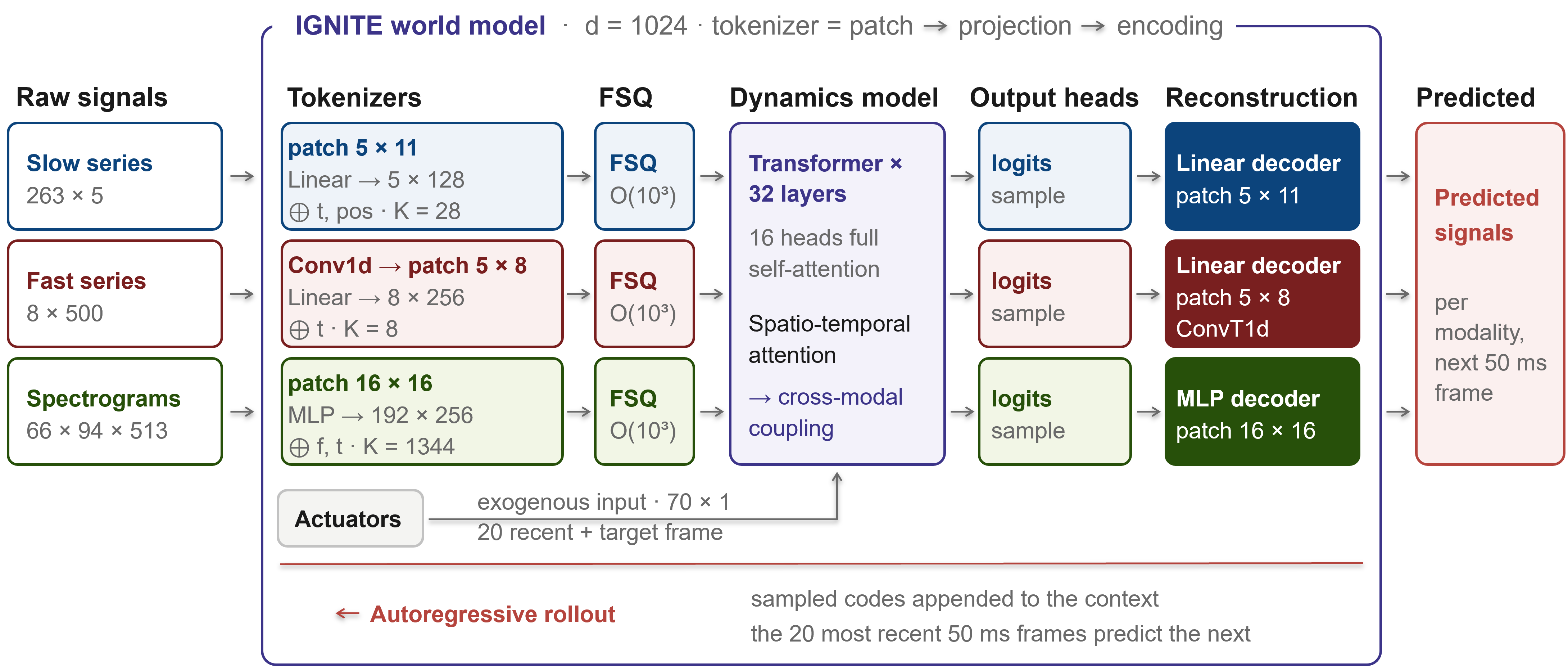}
    \caption{Main outline of the IGNITE world model.}
    \label{fig:IGNITEArchitecture}
\end{figure}

As shown in Figure \ref{fig:IGNITEArchitecture}, IGNITE contains two key components:

\begin{enumerate}
    \item modality-specific tokenizers that convert raw measurements into discrete tokens,
    \item a dynamics model that, given an actuator trajectory and past frame tokens, predicts the tokens of the next frame.
\end{enumerate}

The model is trained in two phases: First, the tokenizers that are used for the dynamics model are trained for encoding and decoding diagnostics. The actuator embeddings and the dynamics model are then trained together for autoregressive rollout.

\subsection{Per-modality tokenization}
\label{subsec:PerModalityTokenization}

Typical world models like \cite{Bruce2024genie,nvidia2025cosmos,Ha2018recurrent,Maes2026leworldmodel} work with uniformly sampled video data or very few modalities that have a common sampling frequency. To accurately simulate a \ddd discharge, however, a caveat is to deal with sampling frequencies that differ in the order of magnitudes, spanning from \SI{100}{\Hz} to \SI{1}{\MHz}. Furthermore, the model has to handle different modalities like multi-channel time-series with low and high sampling frequencies, spectrograms, and further modalities like image sequences. As a result, modality-specific tokenizers are designed to deal with the unique properties of each modality.

In general, we compress each diagnostic frame-based into a discrete token space using \gls{fsq} encoders \cite{Mentzer2024finite}. Each tokenizer takes diagnostics from a window of \SI{50}{\ms} and encodes them into carefully tuned \gls{fsq} encoders. The hyper-parameters and the loss functions for each specific modality are described further in this section. In this work, the modality specific encoders were crucial to process the diagnostics data of wildly different scales. The discrete tokens allow us to reduce the dimensionality and enable a high quality spectrogram generation even in high-frequency ranges. An overview of the exact number of channels, measured properties, and sampling frequencies of all diagnostics can be found in the appended Tables \ref{tab:Diagnostics} and \ref{tab:Actuators}.

\paragraph{Slow time series encoder}

Several diagnostics are an array of probes in the \ddd device which provide spatially rich information about the measurements in different locations of plasma. \gls{ts} provides electron density and temperature at the core and tangential areas of the tokamak cross section \cite{Carlstrom2018thomson}. From \gls{cer}, we use ion temperature and plasma rotation \cite{Chrystal2021charge,Haskey2022details} of the tangential subsystem. We also use measurements of the magnetic pitch angles as captured by the \gls{mse} diagnostic \cite{Holcomb2006motional,Holcomb2008overview}. In total we have seven different categories of slow measurements, each supplied to an individual tokenizer.

Each slow time series has been resampled to \SI{100}{\Hz}. Since IGNITE works with windows of \SI{50}{\ms} duration, each window contains the measurements of all channels for five consecutive samples, i.e., a matrix of number of channels by \num{5}. We refer to Table \ref{tab:Diagnostics} for the number of channel as it depends on the diagnostic. Each tokenizer maps the value of a window to four tokens, each having a dimension of \num{128}. First, the samples in each window are patchified into non-overlapping patches, each carrying spatio-temporal information. The patches are then flattened to a window and linearly projected to \num{128}. Two learned positional embeddings for spatial and temporal encoding are added. Finally, a transformer with four layers and four heads is used to learn spatio-temporal dependencies in the tokens. The resulting tokens of size $4\times\num{128}$ are used as continuous features handed to the \gls{fsq} quantizer.

\paragraph{Fast time series encoder}

While the diagnostics mentioned in the previous paragraph provide spatio-temporal information in slower time pace, other signals like $D_\alpha$ filterscopes capture significantly more temporal information. These signals are sampled at \SI{10}{\kHz} and capture fast transient events like \gls{elm}. These occur as spikes in $D_\alpha$ signals \cite{Colchin2003filterscope}. In this paper, we use the eight filterscope signals that are most commonly used for \ddd experiments.

Each window of \SI{50}{\ms} contains now eight channels each having \num{500} samples. Similar to slow time series, we patchify the signals first, here only in the temporal direction as the amount of spatial information is limited for $D_\alpha$ filterscopes. The patches of all eight channels belonging to the same temporal position are concatenated and represent one token. So, one token carries all the information of eight channels in a window of \SI{10}{\ms} duration. We found that it is important to have a much more expressive latent feature space for fast time series. As a consequence, a solution would be to linearly project the tokens to \num{256} dimensions and then add a learned per-patch positional embedding for learning temporal dependencies. Finally, a transformer with six layers and eight heads is used to learn spatio-temporal dependencies in the tokens. The resulting tokens of size $4\times\num{256}$ are used as continuous features handed to the \gls{fsq} quantizer.

\paragraph{Spectrogram encoder}

The \ddd tokamak has several high-resolution diagnostics that are typically analyzed in the spectral domain. A few examples include \gls{bes} \cite{McKee1999beam} for fluctuation analysis, Bi-color \gls{co2} \cite{VanZeeland2004phase} for density measurements, \gls{ece} radiometer \cite{Austin2003electron} for electron temperature analysis, and \gls{mhr} \cite{Strait2006magnetic} for equilibrium reconstruction, plasma shape control and mode analysis. 

In this paper we resampled all diagnostics to \SI{500}{\kHz} before model training to simplify the data unification and because most of the mode activities can be captured in a frequency range of \SIrange{0}{250}{\kHz}. We computed spectral features using a window size of \num{1024} samples, a hop size of \num{256} and a \textsc{Hann} window. And we disregard the constant DC offset bin in order to get a spectrogram height of \num{512} bins. We use a logarithmic amplitude $\log_{10}(\cdot + \num{1e-5})$ and standardize the resulting spectrograms per frequency bin.

Each window of \SI{50}{\ms} contains \num{98} frames. We patchify the spectrograms in non-overlapping patches of $\num{16}\times\num{16}$ over time and frequency, and we stack the patches of all channels. This leads to differently sized feature vectors that are tokenized subsequently, but to consistently \num{192} tokens per spectrogram modality. The small patch size is a compromise between limiting the number of tokens required for spectrograms and frequency resolution, required for an accurate mode reconstruction. After a linear projection to \num{256} dimension of all tokens, we add trainable positional encodings for time and frequency to the tokens, and use again a transformer with six layers and eight heads to learn spatio-temporal relationships between the tokens. The resulting tokens are again handed to a \gls{fsq} quantizer.

\paragraph{Output heads}

The transformer backbone outputs the predicted token sequence, which contains all tokens for diagnostics in the same order as the input. To decode the diagnostics at the next time step, we thus split the tokens and pass all tokens belonging to a particular diagnostic to a separate decoder that is responsible for reconstructing the actual diagnostic from the predicted token. The decoders are designed symmetrically to the encoders. For slow time series we thus use a linear projection layer, and all convolutional operations are reversed by using transposed convolutions.

Since the autoregressive roll-out takes place solely in the token space, it is not necessary to decode every time step. Instead, the tokens cycle through the backbone during autoregressive roll-out. Decoding is only necessary to compare against the ground truth. During autoregressive model training, we thus use the decoders only for comparing a particular roll-out step with the ground truth when we want to calculate a loss.

\subsection{Shared Transformer Backbone}

We adapt the dynamics model from Genie to be used in IGNITE. It is a decoder-only \gls{maskgit} transformer that processes \gls{fsq} codes. The backbone is a factorized space-time transformer as in \cite{Xu2021spatial}. Since the dynamics model operates purely in the discrete code space, it is effectively a \enquote{language model over plasma-state tokens}. At each time step, it receives the tokens of the different modalities. IGNITE works with a context of \num{20} time windows. The dynamics model thus receives tokens of \num{20} frames. Since the tokens have different dimensions, as described in Section \ref{subsec:PerModalityTokenization}, the first step is to embed them to the model dimension of the dynamics model (\num{1024}) with a linear projection. The actuator signals of each window are passed separately to the backbone. In each window of \SI{50}{\ms}, we first of all calculate the mean value per channel and per actuator signal to obtain a vector of \num{88} dimensions. This is again linearly embedded to the model dimension, and then 

Each of the transformer blocks first calculates cross-attention across all tokens within its own frame for cross-modality mixing, followed by causal temporal self-attention. In this case, each token position attends itself at earlier frames, and this allows information to move through time. Importantly, information can only be passed forward. The output of the dynamic model are then the codes for the next window of \SI{50}{\ms}. To simulate a full \ddd discharge, the output is then appended to the context, and the prediction process repeats until the end of a discharge, simulating the discharge in an autoregressive manner. The predicted codes are finally decoded back into the original modalities (time-series, spectrograms, camera frames) via the frozen decoders, for a comparison against the actual measurements. The dynamics model has \num{1024} dimensions and \num{48} transformer layers, which is a trade-off between training time and accuracy.

\subsection{Training methods}

IGNITE has a two-stage training process: First, we pre-train the codecs for each modality. Afterwards, the codecs are fixed, and the dynamics model is trained to autoregressively predict the full flat-top evolution of a discharge. The training stages are describe subsequently.

\paragraph{Tokenizer pre-training}

The first step of the whole IGNITE training pipeline is to train the codebooks for the modalities. This is considered to be a pre-training step, during which the diagnostic encoders and decoders are trained in order to represent diagnostics by a \gls{fsq}-based codebook. For this purpose, each window of \SI{50}{\ms} duration is treated independently.

\paragraph{Dynamics model training}

The dynamics model closely follows the dynamics model in Genie \cite{Bruce2024genie}. After encoding each modality into integer tokens, each modality $m\in {1,\dots,M}$ ($M=15$) contributes a specific amount of tokens per frame from its codebook of size $V_m$.

We use the \gls{maskgit} approach for training the dynamics model. A training sample consists of \num{100} consecutive frames, each of which being encoded into integer codes. In each window, a randomly selected amount of tokens is masked out, with the masking probability randomize over windows. In this way, some windows have to be predicted completely, whereas other windows are nearly complete. The actuator states in each window are projected to the target dimension (\num{1024}), and added to the corresponding token window. Three more embeddings for modality, positional and temporal encoding are added to each modality. Afterwards, the different modalities are concatenated to form the input to the transformer.

The backbone now consists of \num{24} spatio-temporal transformer blocks. Each block has a pre-normalization layer and a feed-forward network. As in Genie, we use causal masking in the temporal attention, ensuring that only historical information can attend to a given time frame.

The training objective is masked-token prediction. The loss function is the cross-entropy between the class of the target token codec and the predicted codec in the masked sections. The masking and hence the cross entropy is done independently per diagnostic. As a consequence, the final loss is the mean of per-modality losses.

We use the AdamW optimizer \cite{Loshchilov2019decoupled} for weight optimization with weight decay, $\beta=(\num{0.9}, \beta_2)$, and a warmup-cosine schedule, adapted from Genie. Since we are parallelizing over multiple ranks (eight \gls{gpu} nodes, each having eight \gls{gpu}s), the gradients are averaged across ranks by explicit all-reduce.

The dataset suffers from a lot of missing diagnostics. In the case of absent diagnostics during the training process, these are still passed through the model, but the contribution of these diagnostics to the loss function is zero weighted. This is necessary to still allow a modality to receive a gradient signal, which would lead to detection of unused parameters. 
 
During inference, the first \num{20} real frames are rollout seeds. Afterwards, the dynamics model generates each next frame from all-masked in $K$ confidence-ordered unmasking steps, keeping a cosine-decaying masked fraction, sampling the most confident tokens per modality. The predicted codes are appended to the input sequence, and the process repeats.

\section{Dataset}

The IGNITE foundation model was trained on a large dataset of experimental data of the \ddd tokamak, which has one of the most sophisticated sets of diagnostics in the world. The subset of diagnostics used for the experiments in this paper is summarized in Table \ref{tab:Diagnostics}, and the actuators are summarized in Table \ref{tab:Actuators}.

For preparing the original dataset, we collected data from all \ddd discharges between \num{2018} and \num{2026}. Based on the plasma current, we then filtered out successful runs by eliminating disrupted shots where the correlation between the planned plasma current and the actual plasma current falls below a correlation of \num{0.8}. As a result, we get \num{8600} discharges. For model training and evaluation, we randomly split the discharges in \SI{90}{\percent} training, \SI{5}{\percent} validation, and \SI{5}{\percent} test data.

Based on the training dataset, we calculated coefficients for normalization. For each diagnostic, we subtracted the global channel mean and divided by the standard deviation. Since we were dealing with a substantial amount of missing data, we masked these with zeros and excluded these samples from the training loss via the binary mask $M^{(d)}_{k}$. We skipped the first second of each shot, because this initial phase is dominated by a lot of transient effects. We divided the remaining parts of each discharge into non-overlapping chunks of \SI{50}{\ms}, which were used for model training.

\section{Results}
\label{sec:Results}

During model development, we started with overfitting the IGNITE architecture at different scaling levels based on \num{16} shots that belonged to a single experiment. We observed that we needed a significantly higher learning rate of \num{1e-3} compared to Genie (\num{1e-4}), and that we had to train the model for up to \num{80} epochs in order to memorize the training set.

\subsection{Learning curves}
\label{subsec:LearningCurves}

\begin{figure}[!htb]
    \centering
    \subfloat[][\num{1000} shots.]{\includegraphics{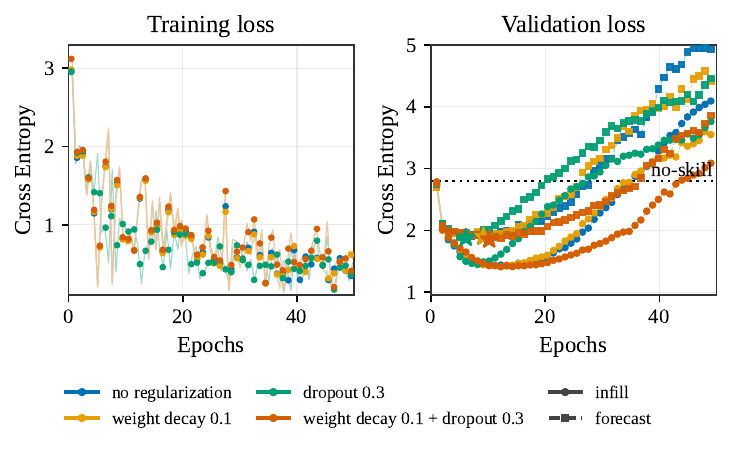}\label{subfig:n1000LearningCurves}}


    \caption{IGNITE learning curves for different regularization levels and dataset sizes. Figure \ref{subfig:n1000LearningCurves}: A subset of \num{1000} shots leads to overfitting, despite regularization. 
    }
    \label{fig:LearningCurves}
\end{figure}

Next, we increased the dataset size to \num{1000} training shots, randomly drawn from the entire training dataset, and trained IGNITE models with weight decay and dropout. The learning curves of training these models for \num{50} epochs are visualized in Figure \ref{subfig:n1000LearningCurves}, and they show that IGNITE tends to overfit on the training dataset strongly. Dropout regularization can slow down the divergence between training and validation dataset. However, the validation loss still started to increase between \num{12} and \num{15} epochs. The training loss was not affected by regularization.

Next, we used the complete training dataset to investigate the impact of dataset size. Following Figure \ref{subfig:n1000LearningCurves}, the minimal validation loss was hardly affected by regularization. Hence, we decided to keep dropout at \num{0.3} and weight decay at \num{0.1} to still enable light regularization. 
Due to the model size, it becomes clear that, despite the model being capable of overfitting to very few shots, generalization requires significantly more data.  Even in an early training stage, the model performance on the validation dataset can now exceed the model trained on \num{1000} shots.

\begin{figure}[!htb]
    \centering
    \includegraphics[width=.8\textwidth,trim={0 180 0 0}, clip]{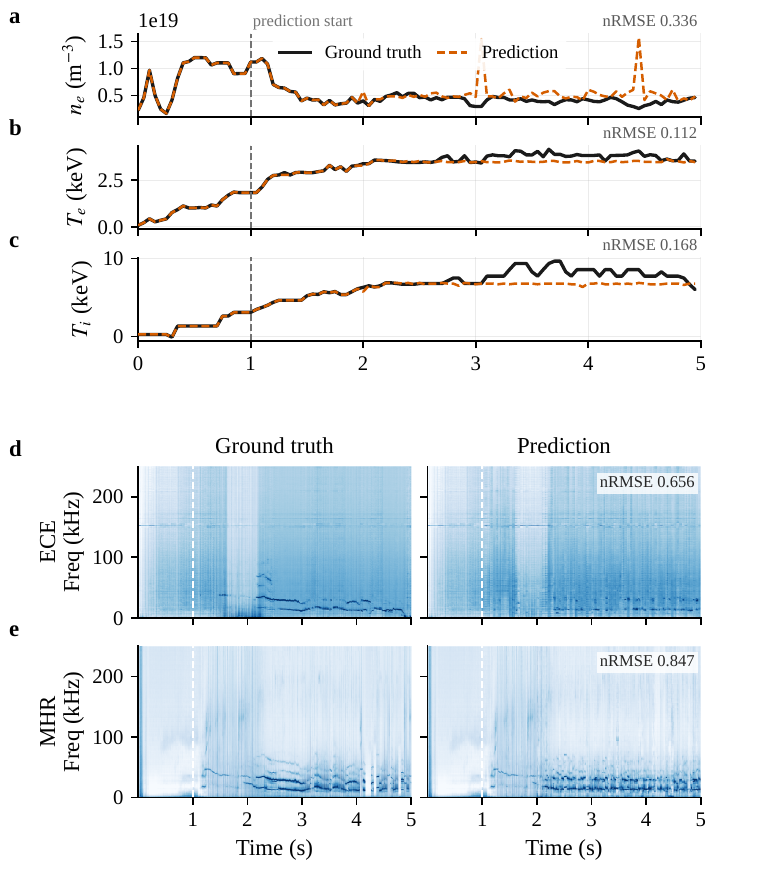}
    \caption{IGNITE predicts the \ddd discharge \num{190735} from the held-out validation set.}
    \label{fig:190735eval}
\end{figure}

\subsection{Example validation output}
\label{subsec:ExampleValidationResult}

This evaluation is based on a representative example from the validation set, namely \ddd \num{190735}. It was not used for any kind of training process, only for monitoring the validation loss across the training progress. Since we did not use early stopping based on validation, we claim that we have not used this shot for any model optimization. From Figure \ref{fig:190735eval} can be observed that IGNITE can predict the full shot trajectory of different modalities given the first second of the shot (\num{20} frames) as an initial seed plus a given actuator trajectory. The spectrograms still suffer from fine-grained mode structures. Especially for \gls{ece}, it becomes clear that the model still has difficulties predicting mode structures in higher frequency ranges. But overall, the prediction accuracy is already at a satisfying level.

\section{Conclusions and future work}
\label{sec:ConclusionsAndFutureWork}

We introduced IGNITE -- a world model for simulating \ddd discharges that has been trained on historical experimental data from the \ddd nuclear fusion device. We showed that a sufficiently large training dataset is crucial for generalization. IGNITE can autoregressively predict signals across a broad variety of modalities.

During model development, we faced several challenges that make our task different to the original Genie architecture. Genie had a training dataset of more than \SI{200000}{\hour} of video clips. After filtering, they ended up with \num{6.8e6} video clips with \SI{16}{\s} duration. For IGNITE, we used \num{8600} sequences of \SI{6}{\s} duration for model training. While Genie is a unimodal world model, ingesting only video data at a constant frame rate, IGNITE ingests \num{14} different modalities in \num{4} different groups, each having a different sampling rate. So, we are dealing with a multi-rate problem. While video data is recorded in a clean environment, \ddd experimental data is collected in a dynamical environment that is impacted by external factors. Diagnostics and actuators can be malfunctioning or not working as programmed. The data quality can suffer from insufficient calibration or noise. And the diagnostic system was growing historically. Channels were added over time, making it difficult to simply use historical data.

Future work includes a physics-based evaluation of this model. We want this to become a useful tool for \ddd researchers to design experiments, to simulate actuator defects that can happen during experiments. This requires us to study the impact of bad/unavailable inputs during training and inference.

\begin{ack}
The authors gratefully acknowledge the collaboration of the \ddd Team supported by the U.S. Department of Energy, Office of Science, Office of Fusion Energy Sciences, using the National Fusion Facility, a \gls{doe} Office of Science user facility, under Award No. DE-FC02-04ER54698, DE-AC02-09CH11466, DE-SC0024527, DE-SC0026429. Additional support was provided by the Princeton Laboratory for Artificial Intelligence under Award No. 2025-97.

This research used resources of the Oak Ridge Leadership Computing Facility at the Oak Ridge National Laboratory, which is supported by the Advanced Scientific Computing Research programs in the Office of Science of the U.S. Department of Energy under Contract No. DE-AC05-00OR22725. 

Disclaimer: This report was prepared as an account of work sponsored by an agency of the United States Government. Neither the United States Government nor any agency thereof, nor any of their employees, makes any warranty, express or implied, or assumes any legal liability or responsibility for the accuracy, completeness, or usefulness of any information, apparatus, product, or process disclosed, or represents that its use would not infringe privately owned rights. Reference herein to any specific commercial product, process, or service by trade name, trademark, manufacturer, or otherwise does not necessarily constitute or imply its endorsement, recommendation, or favoring by the United States Government or any agency thereof. The views and opinions of authors expressed herein do not necessarily state or reflect those of the United States Government or any agency thereof.
\end{ack}


\printbibliography

\clearpage
\appendix
\section*{NeurIPS Paper Checklist}

The checklist is designed to encourage best practices for responsible machine learning research, addressing issues of reproducibility, transparency, research ethics, and societal impact. Do not remove the checklist: {\bf The papers not including the checklist will be desk rejected.} The checklist should follow the references and follow the (optional) supplemental material.  The checklist does NOT count towards the page
limit. 

Please read the checklist guidelines carefully for information on how to answer these questions. For each question in the checklist:
\begin{itemize}
    \item You should answer \answerYes{}, \answerNo{}, or \answerNA{}.
    \item \answerNA{} means either that the question is Not Applicable for that particular paper or the relevant information is Not Available.
    \item Please provide a short (1--2 sentence) justification right after your answer (even for \answerNA). 
\end{itemize}

{\bf The checklist answers are an integral part of your paper submission.} They are visible to the reviewers, area chairs, senior area chairs, and ethics reviewers. You will also be asked to include it (after eventual revisions) with the final version of your paper, and its final version will be published with the paper.

The reviewers of your paper will be asked to use the checklist as one of the factors in their evaluation. While \answerYes{} is generally preferable to \answerNo{}, it is perfectly acceptable to answer \answerNo{} provided a proper justification is given (e.g., error bars are not reported because it would be too computationally expensive'' or ``we were unable to find the license for the dataset we used''). In general, answering \answerNo{} or \answerNA{} is not grounds for rejection. While the questions are phrased in a binary way, we acknowledge that the true answer is often more nuanced, so please just use your best judgment and write a justification to elaborate. All supporting evidence can appear either in the main paper or the supplemental material, provided in appendix. If you answer \answerYes{} to a question, in the justification please point to the section(s) where related material for the question can be found.

IMPORTANT, please:
\begin{itemize}
    \item {\bf Delete this instruction block, but keep the section heading ``NeurIPS Paper Checklist"},
    \item  {\bf Keep the checklist subsection headings, questions/answers and guidelines below.}
    \item {\bf Do not modify the questions and only use the provided macros for your answers}.
\end{itemize}


\begin{enumerate}

\item {\bf Claims}
    \item[] Question: Do the main claims made in the abstract and introduction accurately reflect the paper's contributions and scope?
    \item[] Answer: \answerYes{} 
    \item[] Justification: We have trained a world model for predicting \ddd discharges. In the abstract and introduction, we claim that we can predict the temporal evolution for a given sequence, based on an initial state and an actuator trajectory. We experimental validate this with objective nRMSE measures, by showing the learning curves of the model, and discussing limitations such as generalization and overfitting.

\item {\bf Limitations}
    \item[] Question: Does the paper discuss the limitations of the work performed by the authors?
    \item[] Answer: \answerYes{} 
    \item[] Justification: We discuss that this world model is focused on the \ddd nuclear fusion device and we do not claim that this can be transferred to a different machine. We also do not claim that we can generate physically validated diagnostic evolutions when changing the actuator trajectories. We explicitly state that this is part of the future work.

\item {\bf Theory assumptions and proofs}
    \item[] Question: For each theoretical result, does the paper provide the full set of assumptions and a complete (and correct) proof?
    \item[] Answer: \answerNA{} 
    \item[] Justification: This is an applied \gls{ai} paper. The results are practical and all originate from experiments to optimize the IGNITE model.

    \item {\bf Experimental result reproducibility}
    \item[] Question: Does the paper fully disclose all the information needed to reproduce the main experimental results of the paper to the extent that it affects the main claims and/or conclusions of the paper (regardless of whether the code and data are provided or not)?
    \item[] Answer: \answerYes{} 
    \item[] Justification: All hyper-parameters and implementation details are provided in order to reproduce all results. The code is publicly available, data is available upon request according to \ddd guidelines.

\item {\bf Open access to data and code}
    \item[] Question: Does the paper provide open access to the data and code, with sufficient instructions to faithfully reproduce the main experimental results, as described in supplemental material?
    \item[] Answer: \answerYes{} 
    \item[] Justification: The code is publicly available in our Github repository \href{https://github.com/PlasmaControl/FusionAIHub}{https://github.com/PlasmaControl/FusionAIHub}. We cannot share the data publicly. Following our description to get access to \ddd data, it is possible to preprocess the data with our tools and reproduce all results.

\item {\bf Experimental setting/details}
    \item[] Question: Does the paper specify all the training and test details (e.g., data splits, hyperparameters, how they were chosen, type of optimizer) necessary to understand the results?
    \item[] Answer: \answerYes{} 
    \item[] Justification: We describe all the experiments that lead to the final hyper-parameters. Everything can be reproduced, and we provide all optimizers and processing scripts.

\item {\bf Experiment statistical significance}
    \item[] Question: Does the paper report error bars suitably and correctly defined or other appropriate information about the statistical significance of the experiments?
    \item[] Answer: \answerNo{} 
    \item[] Justification: While we provide objective measurements of the model, we did not run uncertainty quanitification. And we were not able to train several billion parameter models in parallel due to limited computational power.

\item {\bf Experiments compute resources}
    \item[] Question: For each experiment, does the paper provide sufficient information on the computer resources (type of compute workers, memory, time of execution) needed to reproduce the experiments?
    \item[] Answer: \answerYes{} 
    \item[] Justification: We describe which hardware we used and how many resources. We also describe the training and inference time. 
    
\item {\bf Code of ethics}
    \item[] Question: Does the research conducted in the paper conform, in every respect, with the NeurIPS Code of Ethics \url{https://neurips.cc/public/EthicsGuidelines}?
    \item[] Answer: \answerYes{} 
    \item[] Justification: We do not use any personal data for the studies in this paper.

\item {\bf Broader impacts}
    \item[] Question: Does the paper discuss both potential positive societal impacts and negative societal impacts of the work performed?
    \item[] Answer: \answerNA{} 
    \item[] Justification: IGNITE does not use any kind of private data. It can potentially be used to generate unexpected diagnostic trajectories. However, since every \ddd-related publication undergoes an internal review process, physicists will easily recognize if unreasonable outputs were generated with IGNITE.
    
\item {\bf Safeguards}
    \item[] Question: Does the paper describe safeguards that have been put in place for responsible release of data or models that have a high risk for misuse (e.g., pre-trained language models, image generators, or scraped datasets)?
    \item[] Answer: \answerNA{} 
    \item[] Justification: We are not allowed to release the data used to train IGNITE without users signing user agreements with \ddd. So, there is no safety risk as we will not release the dataset. Model checkpoints will become available upon reasonable request.

\item {\bf Licenses for existing assets}
    \item[] Question: Are the creators or original owners of assets (e.g., code, data, models), used in the paper, properly credited and are the license and terms of use explicitly mentioned and properly respected?
    \item[] Answer: \answerYes{} 
    \item[] Justification: We are the code owners, and the code is in our Github repository. We used Claude Code for software development. We cite all utilized foundations for model development. We acknowledge the cluster that we used for model training. We acknowledge the \ddd team for allowing us to collect and use the dataset.

\item {\bf New assets}
    \item[] Question: Are new assets introduced in the paper well documented and is the documentation provided alongside the assets?
    \item[] Answer: \answerYes{} 
    \item[] Justification: We describe the training and validation process of the IGNITE model in details, and the limitations of the model. License is in the FAITH repository.

\item {\bf Crowdsourcing and research with human subjects}
    \item[] Question: For crowdsourcing experiments and research with human subjects, does the paper include the full text of instructions given to participants and screenshots, if applicable, as well as details about compensation (if any)? 
    \item[] Answer: \answerNA{} 
    \item[] Justification: Our research did not involve crowdsourcing or research with human subjects. 

\item {\bf Institutional review board (IRB) approvals or equivalent for research with human subjects}
    \item[] Question: Does the paper describe potential risks incurred by study participants, whether such risks were disclosed to the subjects, and whether Institutional Review Board (IRB) approvals (or an equivalent approval/review based on the requirements of your country or institution) were obtained?
    \item[] Answer: \answerNA{} 
    \item[] Justification: Our research did not involve crowdsourcing or research with human subjects. 

\item {\bf Declaration of LLM usage}
    \item[] Question: Does the paper describe the usage of LLMs if it is an important, original, or non-standard component of the core methods in this research? Note that if the LLM is used only for writing, editing, or formatting purposes and does \emph{not} impact the core methodology, scientific rigor, or originality of the research, declaration is not required.
    \item[] Answer: \answerNo{} 
    \item[] Justification: We designed and developed the model based on the Genie paper \cite{Bruce2024genie}. Every development step was decided by us. We used Claude Code to assist with the implementation. However, we carefully reviewed the code and double-checked the results. We used unit-tests to guide the software development properly. For the manuscript, we used AI tools to make the main figure more concise, after preparing the initial version.

\end{enumerate}

\clearpage

\begin{landscape}
\section{Utilized diagnostics and actuators}

Table \ref{tab:Diagnostics} and Table \ref{tab:Actuators} summarize the diagnostics and actuators that were used for training the IGNITE world model. We grouped the diagnostics as specified in Section \ref{subsec:PerModalityTokenization} and resampled the diagnostics in each group. 

\begin{table}[!htb]
    \renewcommand{\arraystretch}{1.3}
    \centering
    \caption{Overview of the diagnostics used for this study.}
    \label{tab:Diagnostics}
    \begin{tabular}{lllrrl}
        \toprule
        Diagnostic & Measured properties & Channels & Sampling freq. & Resampled freq. & Modality type \\
        \midrule
        \gls{cer} & Ion temperature ($T_\mathrm{i}$), plasma rotation ($v_\mathrm{tor}$) & $\num{2}\times\num{48}$ & \SI{100}{\Hz} & -- & Slow time series \\
        \gls{mse} & Magnetic pitch angle & \num{69} & \SI{100}{\Hz} & -- & Slow time series \\
        \gls{ts} & Electron temperature ($T_\mathrm{e}$) and density ($n_\mathrm{e}$) & $\num{2}\times\num{44}$ \& $\num{2}\times\num{10}$ & \SI{100}{\Hz} & -- & Slow time series \\
        \midrule
        Filterscopes & Line emissions & \num{8} & \SI{10}{\kHz} & -- & Fast time series \\
        \midrule
        \gls{bes} & Fluctuations & \num{16} & \SI{1}{\MHz} & \SI{500}{\kHz} & Spectrogram \\
        \gls{ece} & Electron temperature & \num{48} & \SI{500}{\kHz} & -- & Spectrogram \\
        \gls{co2} & Line-averaged electron density & \num{4} & \SI{1.3}{\MHz} & \SI{500}{\kHz} & Spectrogram \\
        Magnetics & Magnetic field & \num{6} & \SI{1}{\MHz} & \SI{500}{\kHz} & Spectrogram \\
        \bottomrule
    \end{tabular}
\end{table}

\begin{table}[!htb]
    \renewcommand{\arraystretch}{1.3}
    \centering
    \caption{Overview of the actuators used for this study.}
    \label{tab:Actuators}
    \begin{tabular}{llrr}
        \toprule
        Actuator & Actuated property & Channels & Sampling frequency \\
        \midrule
        \gls{ech} & Power, toroidal angle, poloidal angle, polarization & \num{8} & \SI{10}{\kHz} \\
        Gas & Voltage, gas flow & \num{8} & \SI{10}{\kHz} \\
        \gls{nbi} & Power, torque, voltage & \num{8} & \SI{10}{\kHz} \\
        \gls{rmp} & Coil current & \num{8} & \SI{10}{\kHz} \\
        I-Coil & Coil current & \num{12} & \SI{10}{\kHz} \\
        \bottomrule
    \end{tabular}
\end{table}

\end{landscape}

\section{Details on the modality-specific tokenizers}

\begin{figure}[!htb]
    \centering
    \includegraphics[width=\textwidth]{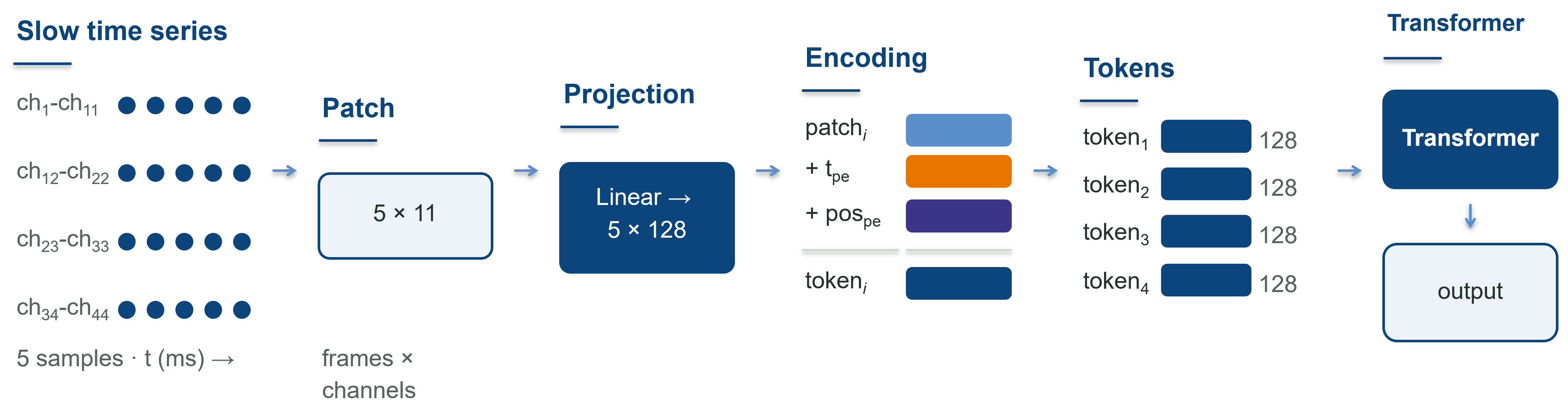}
    \caption{Tokenizer for slow time series. The channel-time array is grouped into four channel groups and patched at 5 samples $\times$ channels; a shared linear layer projects each patch by Linear $\rightarrow 5\times\num{128}$; a time encoding and a sequence-position encoding are added to each token; the token sequence, each 128-dimensional, is processed by the transformer to produce the model output.}
    \label{fig:SlowTSTokenizer}
\end{figure}

Figure \ref{fig:SlowTSTokenizer} illustrates the processing pipeline for slow time-series.

\begin{figure}[!htb]
    \centering
    \includegraphics[width=\textwidth]{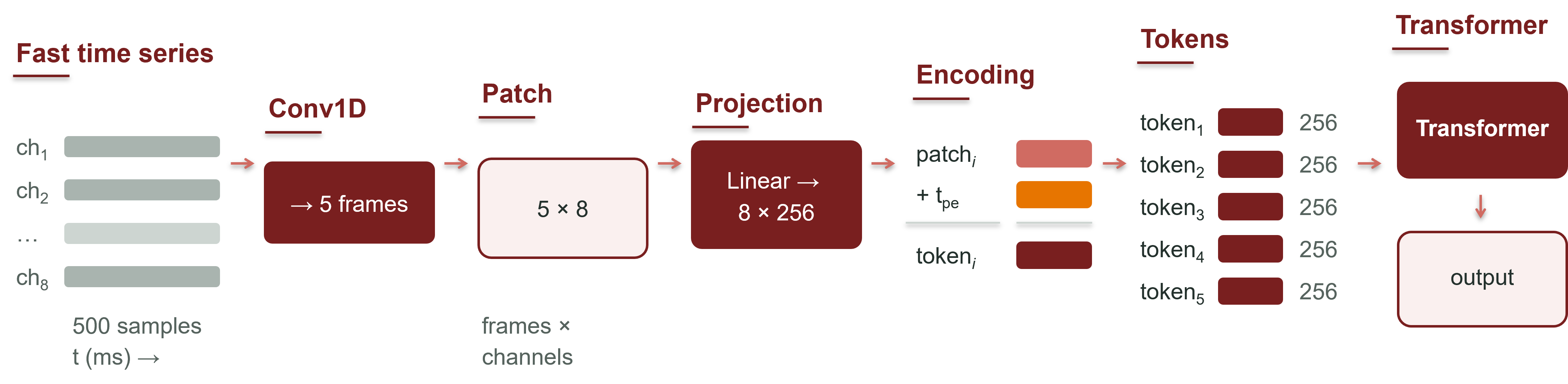}
    \caption{Tokenizer for fast time series. A \SI{50}{\ms} window of 8 channels sampled at \SI{10}{\kHz} gives \num{500} samples per channel; a Conv1D front end reduces each channel to 5 frames, patched at $5 \times 8$ over frames and channels. A shared Linear $\rightarrow 8 \times \num{256}$ projects each patch, a \num{256}-wide time encoding is added, and the token sequence is processed by the transformer to produce the model output.}
    \label{fig:FastTSTokenizer}
\end{figure}

Figure \ref{fig:FastTSTokenizer} illustrates the processing pipeline for fast time-series.

\begin{figure}[!htb]
    \centering
    \includegraphics[width=\textwidth]{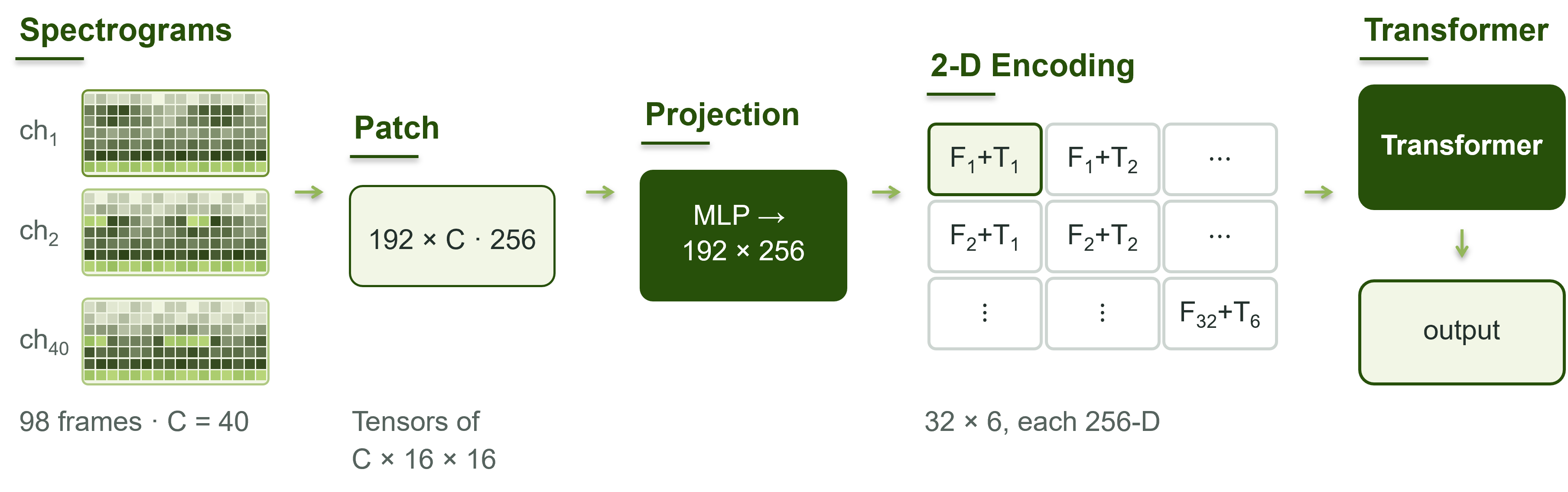}
    \caption{Tokenizer for spectrograms. SEach of the $C$ channels contributes a time-frequency map of 98 frames, cut into $16 \times 16$ patches, giving \num{192} patches per channel. Patches at the same grid position are concatenated across channels and an MLP reduces them to $\num{192} \times \num{256}$. A frequency encoding ($\num{32} \times \num{256}$) and a time encoding ($6 \times \num{256}$) form a $32 \times 6$ table whose entries are 256-dimensional vectors, added to the token at that grid position, before the transformer produces the model output.}
    \label{fig:SpectrogramTokenizer}
\end{figure}

Figure \ref{fig:SpectrogramTokenizer} illustrates the processing pipeline for spectrograms.

\end{document}